\documentclass[final,1p,times]{elsarticle}
\usepackage{hyperref}
\usepackage{amssymb}
\usepackage{amsmath}
\usepackage{color}
\usepackage{soul}
\usepackage{graphicx}
\begin{document}

\begin{frontmatter}

\title{Asymptotic potential of a rose-shaped disk} 

\author[1]{Alina E. Sagaydak}
\author[1,2]{Zurab K. Silagadze}

\affiliation[1]{organization={Novosibirsk State University},
            addressline={}, 
            city={Novosibirsk},
            postcode={630090},
            state={},
            country={Russia}
            }

\affiliation[2]{organization={Budker Institute of Nuclear Physics},
            addressline={}, 
            city={Novosibirsk},
            postcode={630090},
            country={Russia}
            }

\ead{a.sagaidak@g.nsu.ru}
\ead{silagadze@inp.nsk.su}

\begin{abstract}
Based on extensive numerical evidence, a recent paper (Sheng Chen et al., Eur. J. Phys. {\bf 45} (2024), 045703) suggested that the potential of a uniformly charged rose-shaped disk in the plane of the disk has a simple asymptotic form. We present an analytical proof of this interesting conjecture.
\end{abstract}



\end{frontmatter}

\section{Introduction}
The calculation of the equilibrium charge distribution even for regular bodies is one of the most difficult problem in potential theory \cite{Durand_1953,Kondratyev_2007,Griffiths_1996,Good_1996}. For this reason, many authors simplify the problem by assuming a uniform charge distribution \cite{Pollack_2021,Ciftja_2021,Ciftja_2020} that generally is not the equilibrium distribution. Even for classical cases such as a uniformly charged ring \cite{Ciftja_2009} or disk \cite{Bochko_2020}, the calculation of the electrostatic potential created by these bodies at an arbitrary point in space are quite challenging. Fewer results are known for objects with more exotic shapes such as rose-shaped disks or likewise structures. For this reason, in this paper we focus on the rose disk, for whose electrostatic potential an interesting and simple asymptotic form has recently been proposed \cite{Chen_2024}.

The potential of a uniformly charged rose-shaped disk at the observation point with polar coordinates $(R,\theta)$ in the plane of the disk has the form
\begin{eqnarray} && \hspace*{-20mm} 
U(R,\theta)=\frac{\sigma}{4\pi\epsilon_0}\int\limits_0^{\alpha_{max}}d\alpha\int\limits_0^{a\cos{(n\alpha)}}\frac{\rho d\rho}{\sqrt{(\rho\cos{\alpha}-R\cos{\theta})^2+(\rho\sin{\alpha}-R\sin{\theta})^2}}=\nonumber \\ && \hspace*{-20mm}
\frac{\sigma}{4\pi\epsilon_0}\int\limits_0^{\alpha_{max}}d\alpha\int\limits_0^{a\cos{(n\alpha)}}\frac{\rho d\rho}{\sqrt{R^2-2\rho R\cos{(\theta-\alpha)}+\rho^2}},\;\;\alpha_{max}=\left\{\begin{array}{c} 2\pi,\;\;
n\;\mathrm{even},\\ \pi,\;\; n\;\mathrm{odd}. \end{array}\right .
\label{eq1}
\end{eqnarray}
where $\sigma$ denotes the surface-charge density of the rose disk, $\epsilon_0$ is the dielectric constant (permittivity) in vacuum, $a$ is the radius of the circumscribed circle around the rose disk, and $n$ is an integer indicating the number of petals: $2n$ if $n$ is even, and $n$ if $n$ is odd. When $n$ is odd, the petals for $\alpha>\pi$ are in the same place as for $\alpha<\pi$. Therefore, in this case $\alpha_{max}=\pi$ to avoid double counting.

In \cite{Chen_2024}, through extensive numerical testing, the following simple asymptotic formula for the potential (\ref{eq1}) when $R\gg a$ was proposed
\begin{equation}
U(R,\theta)\approx \left\{\begin{array}{lc} A_n(R)\cos{(n\theta)}+B_n(R), & \mathrm{if}\;\; n\;\mathrm{is}\;\mathrm{odd},\\ A_n(R)\cos{(2n\theta)}+B_n(R),& \mathrm{if}\;\; n\;\mathrm{is}\; \mathrm{even}.\end{array}\right .
\label{eq2}
\end{equation}
The purpose of this short note is to provide analytical proof of this conjecture.

\section{Some mathematical preliminaries}
In this section, we will introduce some mathematical background that will help us in our proof. These identities are well known in the mathematical literature, but for the convenience of the readers we provide their proofs in the appendix.

Using \cite{Boros_Moll_2004}
\begin{equation}
\Bigl( \frac{1}{2}\Bigr)_n=\frac{(2n-1)!!}{2^n},\;\;\binom{n}{k}=\frac{n!}{k!(n-k)!},\;\;
(2n-1)!!=\frac{(2n)!}{2^n n!},
\label{eq9}
\end{equation}
we rewrite (\ref{eq8}) in the form of the first identity, which we will need in the following:
\begin{equation}
P_n(\cos{\theta})=\sum_{k=0}^n \frac{1}{2^{2n}}\binom{2k}{k}\binom{2(n-k)}{n-k}\, \cos{(n-2k)\theta}.  
\label{eq10} 
\end{equation}
The second required identity is
\begin{equation}
\cos^n{\theta}=
\frac{1}{2^n}\sum_{k=0}^n \binom{n}{k}\cos{(n-2k)\theta}.
\label{eq11}
\end{equation}

\section{Multipole expansion of the potential}
With $t=\rho/R\ll 1$, equation (\ref{eq3}) implies
\begin{equation}
U(R,\theta)=\frac{\sigma}{4\pi\epsilon_0R}\int\limits_0^{\alpha_{max}}d\alpha\int\limits_0^{a\cos{(n\alpha)}}\rho d\rho\sum_{m=0}^\infty P_m(\cos{(\theta-\alpha)})\,\Bigl (\frac{\rho}{R}\Bigr)^m.
\label{eq12}
\end{equation}
According to the Fubini-Tonelli theorem, if the double integral of the absolute value is finite, then the order of integration does not matter \cite{Weir_1973}, and summation can be considered as integration with the counting measure \cite{Shirali_2018}. As the absolute value of the Legendre polynomial is no larger than one, we have
\begin{equation}
\sum_{m=0}^\infty\int\limits_0^{a\cos{(n\alpha)}}\rho d\rho\left |P_m(\cos{(\theta-\alpha)})\right |\,\Big(\frac{\rho}{R}\Bigr )^m< 
\sum_{m=0}^\infty\int\limits_0^{a\cos{(n\alpha)}}\rho d\rho\,\Bigr(\frac{\rho}{R}\Bigl)^m=\sum_{m=0}^\infty a_m<\infty,
\label{eq13}
\end{equation}
with
\begin{equation}
a_m=\frac{a^{m+2}\cos^{m+2}{(n\alpha)}}{(m+2)R^m}.
\label{eq14}
\end{equation}
The convergence of the last series follows from the d'Alembert-Cauchy ratio criterion, since
\begin{equation}
\lim_{m\to\infty}\left|\frac{a_{m+1}}{a_m}\right |=\frac{a}{R}|\cos{(n\alpha)}|<1.
\label{eq15}
\end{equation}
Therefore, we can swap the second integration and summation in (\ref{eq12}) and get
\begin{equation}
U(R,\theta)=\frac{\sigma}{4\pi\epsilon_0R}\int\limits_0^{\alpha_{max}}d\alpha \sum_{m=0}^\infty P_m(\cos{(\theta-\alpha)})\,a_m.
\label{eq16}
\end{equation}    
We can also swap the remaining integration and summation because again using the d'Alembert-Cauchy ratio criterion and $\left |P_m(\cos{(\theta-\alpha)})\cos^{m+2}{(n\alpha)}\right |<1$ we get
\begin{equation}
\sum_{m=0}^\infty \int\limits_0^{\alpha_{max}}d\alpha \left | P_m(\cos{(\theta-\alpha)})\,a_m\right |<
\sum_{m=0}^\infty\frac{\alpha_{max}a^{2}}{m+2}\,\Bigr(\frac{a}{R}\Bigl)^m<\infty.
\label{eq17}
\end{equation}
Finally,
\begin{equation}
U(R,\theta)=\frac{\sigma a^2}{4\pi\epsilon_0R}\sum_{m=0}^\infty\frac{1}{m+2}\Bigl(\frac{a}{R}\Bigr)^m\int\limits_0^{\alpha_{max}} d\alpha \, P_m(\cos{(\theta-\alpha)})\cos^{m+2}{(n\alpha)}.
\label{eq18}
\end{equation}
To calculate the angular integral
\begin{equation}
I(n,m,\theta)=\int\limits_0^{\alpha_{max}} d\alpha \, P_m(\cos{(\theta-\alpha)})\cos^{m+2}{(n\alpha)},
\label{eq19}
\end{equation}
we use expansions (\ref{eq10}) and (\ref{eq11}) to obtain
\begin{equation}
I(n,m,\theta)=\sum_{s=0}^{m+2}\sum_{l=0}^m A^{(m+2)}_sC^{(m)}_l \int\limits\limits_0^{\alpha_{max}} d\alpha \cos{\left[(m-2l)(\theta-\alpha)\right]}\cos{\left [(m+2-2s)n\alpha\right ]},
\label{eq20}
\end{equation}
where
\begin{equation}
A^{(m)}_s=\frac{1}{2^m}\binom{m}{s},\;\;\; C^{(m)}_l=\frac{1}{2^{2m}} \binom{2l}{l}\binom{2(m-l)}{m-l}.    
\label{eq21}
\end{equation}
We can use the product-to-sum formula for cosines to simplify the integration into (\ref{eq20}):
\begin{eqnarray*} &&
\int\limits_0^{\alpha_{max}} d\alpha \cos{\left[(m-2l)(\theta-\alpha)\right]}\cos{\left [(m+2-2s)n\alpha\right ]}= \nonumber \\ &&
\frac{1}{2}\int\limits_0^{\alpha_{max}} d\alpha \cos{\left\{(m-2l)\theta-[(m-2l)+(m+2-2s)n]\,\alpha\right \}} +\\ &&
\frac{1}{2}\int\limits_0^{\alpha_{max}} d\alpha \cos{\left\{(m-2l)\theta-[(m-2l)-(m+2-2s)n]\,\alpha\right \}}.
\end{eqnarray*}
If the expression preceding the $\alpha$ is not equal to zero and $\alpha_{max} = 2\pi$, then the resulting sine values at both the upper and lower limits will be the same and the integral will be zero. If $\alpha_{max} = \pi$, then $n$ will be odd. Consequently, both $n+1$ and $n-1$ will be even and, if the expression preceding the $\alpha$ is not equal to zero, the integral similarly will again be zero. If the expression preceding the $\alpha$ is equal to zero, integration into (\ref{eq20}) becomes trivial. The final result is
\begin{eqnarray} &&
\int\limits_0^{\alpha_{max}} d\alpha \cos{\left[(m-2l)(\theta-\alpha)\right]}\cos{\left [(m+2-2s)n\alpha\right ]}= \nonumber \\ &&
\cos{\left[ (m+2-2s)n\theta\right ]}\left [\delta_{n(m+2-2s),m-2l}+\delta_{n(m+2-2s),2l-m}\right ]\frac{\alpha_{max}}{2}.
\label{eq22}
\end{eqnarray}
The first Kronecker delta function is non-zero only when
\begin{equation}
2l=2l_1=2n(s-1)-m(n-1),
\label{eq23}
\end{equation}
while the second delta function is non-zero only when
\begin{equation}
2l=2l_2=m(n+1)-2n(s-1).
\label{eq24}
\end{equation}
As we see, both $m(n-1)$ and $m(n+1)=m(n-1)+2m$ must be even numbers for the integral $I(n,m,\theta)$ to be non-zero. Let $m(n-1)=2k$. Then $l_1=n(s-1)-k$ and $l_2=m+k-n(s-1)=m-l_1$. 
But from (\ref{eq21}) $C^{(m)}_l=C^{(m)}_{m-l}$. Therefore, the second Kronecker delta function in (\ref{eq22}) makes exactly the same contribution to $I(n,m,\theta)$ as the first Kronecker delta function.

It remains to refine the summation limits in $s$ for equation (\ref{eq20}). Since $2l_1=2n(s-1)-m(n-1)\ge 0$, we must have $s-1\ge \frac{m(n-1)}{2n}$. Therefore,
\begin{equation}
s\ge 1+\lceil  \frac{m(n-1)}{2n} \rceil,
\label{eq25}    
\end{equation}
where $\lceil x \rceil$ is the ceiling function -- the smallest integer greater than or equal to $x$. On the other hand, $2l_1\le 2m$ and similarly we obtain
\begin{equation}
s\le 1+\lfloor  \frac{m(n+1)}{2n} \rfloor,
\label{eq26}    
\end{equation}
where $\lfloor x \rfloor$ is the floor function -- the greatest integer less than or equal to $x$. 

Finally, we can summarize the multipole expansion of the potential as follows:
\begin{equation}
U(R,\theta)=\frac{\sigma a^2}{4\pi\epsilon_0R}\sum_{m=0}^\infty\frac{1}{m+2}\Bigl(\frac{a}{R}\Bigr)^m I(n,m,\theta),
\label{eq27}
\end{equation}
where $I(n,m,\theta)$ is zero if $m(n-1)$ is an odd number, or $\lceil  \frac{m(n-1)}{2n} \rceil > \lfloor  \frac{m(n+1)}{2n} \rfloor$. In other cases
\begin{equation}
I(n,m,\theta)=\alpha_{max}\sum_{s=1+\lceil  \frac{m(n-1)}{2n} \rceil}^{1+\lfloor  \frac{m(n+1)}{2n} \rfloor}A^{(m+2)}_sC^{(m)}_{n(s-1)-\frac{m(n-1)}{2}} \cos{\left[ (m+2-2s)n\theta\right ]}.
\label{eq28}
\end{equation}

\section{The asymptotic form of the rose-disk potential}
The monopole $m=0$ term (Coulomb potential) is always present. It is proportional to the total charge $Q$ of the disk and makes the main contribution to the constant term in (\ref{eq2}):
\begin{equation}
B_n(R)= \frac{Q}{4\pi\epsilon_0 R},\;\;  Q=\frac{1}{4}\sigma\alpha_{max}a^2.
\label{eq29}
\end{equation}
If $n=1$, the next-to-dominant contribution gives the dipole term $m=1$ and
\begin{equation}
I(1,1,\theta)=\pi(A^{(3)}_1C^{(1)}_0+A^{(3)}_2C^{(1)}_1)\cos{\theta}=\frac{3}{8}\pi\cos{\theta}.
\label{eq30}
\end{equation}
Therefore,
\begin{equation}
A_1(R)=\frac{\sigma a^3}{32\epsilon_0R^2}.
\label{eq31}
\end{equation}
For $n>1$ we need to find the smallest $m$ for which $I(n,m,\theta)$ is nonzero, and this will determine the leading higher order multipole. 

Let $n=2k+1$ and $m=2l+1$. Then the inequality 
\begin{equation}
\frac{m(n-1)}{2n}\le s-1\le \frac{m(n+1)}{2n}
\label{eq32}
\end{equation}
can be rewritten in the following form:
\begin{equation}
l+\frac{k-l}{2k+1}\le s-1\le l+1-\frac{k-l}{2k+1}.
\label{eq33}
\end{equation}
If $m<n$, that is if $l<k$, then $0<\frac{k-l}{2k+1}<\frac{1}{2}$ and it is clear that (\ref{eq33}) does not have an integer solution for $s$. Therefore, for $n$ and $m$ both odd,
the smallest $m$ for which $I(n,m,\theta)$ is not zero is given by $m=n$. In this case, inequality (\ref{eq33}) has two integer solutions: either $s=l+1$ or $s=l+2$. In the first case, $m+2-2s=1$, and in the second, $m+2-2s=-1$. In both cases, the corresponding contribution to the potential is proportional to $\cos{(n\theta)}$.

If $n=2k+1$, but $m=2l$, then the inequality (\ref{eq32}) takes the form
\begin{equation}
l-\frac{l}{2k+1}\le s-1\le l+\frac{l}{2k+1}.
\label{eq34}
\end{equation}
For $0<m<n$, $0<\frac{l}{2k+1}<\frac{1}{2}$, and (\ref{eq34}) has an unique integer solution $s=l+1$. But for this solution $m+2-2s=0$ and, therefore, the corresponding multipoles contribute only to the constant (not dependent on $\theta$) part $B_n(R)$ of (\ref{eq2}). This completes the proof of (\ref{eq2}) when $n$ is odd.

Let now $n=2k$ be even. Only when $m=2l$ is even, the angular integral $I(n,m,\theta)$ is not equal to zero, since for this $m(n-1)$ must be even. The inequality (\ref{eq32}) takes the form
\begin{equation}
l-\frac{l}{2k}\le s-1\le l+\frac{l}{2k}.
\label{eq35}
\end{equation}
If $0<l<2k$, this inequality has only one integer solution $s=l+1$. But for this solution $m+2-2s=0$ and the corresponding multipoles contribute only to the constant part $B_n(R)$ of (\ref{eq2}).

If $l=2k$, that is if $m=2n$, (\ref{eq35}) has three integer solutions: $s=l$, $s=l+1$ or $s=l+2$. In the first case, $m+2-2s=2$ and the corresponding contribution to the potential is proportional to $\cos{(2n\theta)}$. In the second case, $m+2-2s=0$ and thus the corresponding contribu
ion to the potential does not depend on $\theta$. In the third case, $m+2-2s=-2$ and the corresponding contribution to the potential is again proportional to $\cos{(2n\theta})$. Therefore, (\ref{eq2}) is valid also when $n$ is even.

\begin{figure}[htp]
\begin{center}
\includegraphics[scale=1.2]{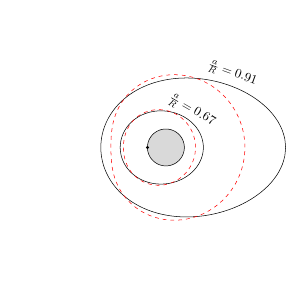}
\includegraphics[scale=1.2]{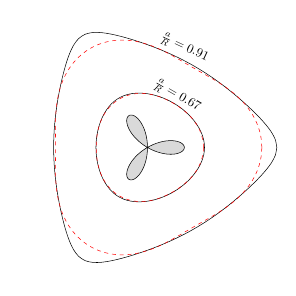}
\includegraphics[scale=1.2]{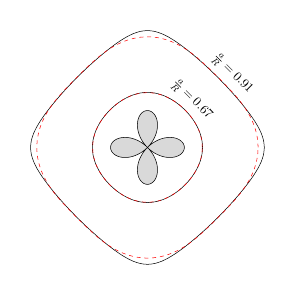}
\includegraphics[scale=1.2]{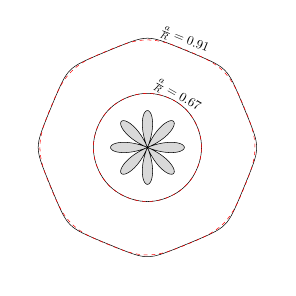}
\end{center}
\caption{The figure illustrates the accuracy of the approximation by multipoles, up to the multipole where the angular dependence first appears. The solid line represents the numerically calculated potential (in units of $u=\frac{\sigma R}{4\pi\epsilon_0}$ and in dimensionless polar coordinates $(U/u,\theta)$) for fixed values of $a/R$. The red dashed line represents the multipole approximation.}
\label{fig1}
\end{figure}

Fig.\ref{fig1} shows the accuracy of the expression (\ref{eq2}) when approximating the potential (\ref{eq1}). The greatest difference is observed in the direction of the petals, which are also shown schematically in the figure. However, already for $a/R=0.5$ the maximum difference is less than one percent for all the cases considered, with the exception of $n=1$, where the maximum difference was about 8\%.

For $n=1$, the rose-shaped disk is actually a circular disk offset from the origin (see Fig.\ref{fig1}). An interesting story is connected with the gravitational analogue of the electrostatic potential of a uniformly charged round disk.

The Voyager space probes used a trajectory with multiple gravity assists, made possible by a configuration of the four outer planets that occurs only once every 175 years. To use a gravity assist from Jupiter, Voyager 2 had to squeeze through a very narrow corridor 150 km wide and arrive at a specific point with an accuracy of one second. If Voyager missed this point or was off by a few seconds, it would ruin the gravity assist and the subsequent gravity assist around Saturn.

Voyager-2 made two more gravity assists around Saturn and Uranus, allowing it to make incredible astronomical discoveries. The last gravity assist was around Neptune to study its moon Triton.

The gravity assist maneuver near Saturn also required great precision in order to later explore Saturn's rings and its satellites. But there was a problem: it was necessary to take into account the gravitational field of Saturn's rings. This led to two publications \cite{Lass_1983, Krogh_1982}, in which the gravitational potential of a homogeneous disk and the corresponding gravitational field were calculated. ``With these results, Voyager-2 passed within 150 kilometers of its target point in the B-plane when it reached Uranus. The B-plane is the plane perpendicular to the spacecraft's trajectory through the center of the planet. Without explicit expressions for the gravitational attraction of Saturn's rings, Voyager-2 could not have been so skillfully navigated toward Uranus and then Neptune." (W. Van Snyder -- private communication).

This story vividly illustrates the unexpected usefulness of even seemingly purely academic exercises in electrostatics, in full accordance with the wisdom of the medieval philosopher, theologian and mystical writer Hugo of Saint-Victor: "Study everything. Later you will see that there is nothing superfluous."

Recently, it was shown in \cite{Sagaydak_2025} that the equivalence of Green's theorem to the divergence theorem in two dimensions allows one to more easily obtain the results of \cite{Lass_1983, Krogh_1982}. In particular, the electrostatic potential created by a uniformly charged flat region $D$ with a surface charge density $\sigma$ at the observation point $(X,Y,Z)$ is determined by the linear integral over the boundary of the body $\partial D$:
\begin{equation}
U(X,Y,Z)=\frac{\sigma}{4\pi\epsilon_0}\oint \frac{(x -X)dy-(y-Y)dx}{\sqrt{(x -X)^2+(y-Y)^2+Z^2}+|Z|}.
\label{eq1DP}
\end{equation}
In our case, the boundary of the rose-shaped disk is given by the equations
$$x=a\cos{(n\alpha)}\cos{\alpha},\;\;y=a\cos{(n\alpha)}\sin{\alpha},$$
and (\ref{eq1DP}) after rather long but simple algebra takes the form
\begin{equation}
U(X,Y,Z)=\frac{\sigma}{4\pi\epsilon_0}\int\limits_0^{\alpha_{max}} \frac{a^2\cos^2{n\alpha}-aR\left[ \cos{n\alpha}\cos{(\theta-\alpha)}+n\sin{n\alpha}\sin{(\theta-\alpha)}\right]}{\sqrt{R^2+Z^2+a^2\cos^2{n\alpha}-2aR\cos{n\alpha}\cos{(\theta-\alpha)}}+|Z|}\,d\alpha     
\label{eq1DPRD}
\end{equation}
This formula allows us to calculate the potential not only in the plane of the disk, but also at any point in space.

\section{Concluding remarks}
Thus we have proved that the conjecture (\ref{eq2}) about the asymptotic form of the rose-disk potential, suggested in \cite{Chen_2024}, is valid. Another interesting property of the rose-disk potential is its low sensitivity to the number of petals \cite{Aghamohammadi_2024}. More precisely, the electric potentials at a given point $(0,0,Z)$ on the axis of symmetry are the same for all even $n$, and the electric potentials for all odd $n$ are also the same. However, since $\alpha_{max}$ differs by a factor of two for even and odd $n$, the corresponding electric potentials also differ by a factor of two. The formula for electric potential is
\begin{equation}
U(0,0,Z)=\frac{\sigma\alpha_{max}}{4\pi\epsilon_0}\left [\frac{2\sqrt{a^2+Z^2}}{\pi}\,E(k)-|Z|\right ],
\label{eq36}
\end{equation}
where
\begin{equation}
    k=\sqrt{\frac{a^2}{a^2+z^2}},
\label{eq37}    
\end{equation}
and $E(k)$ is the complete elliptic integral of the second kind. This formula was obtained in \cite{Aghamohammadi_2024} for a petal-shaped disk given by $r=a|\cos{n\alpha}|$. For a rose-shaped disk $x=a\cos{(n\alpha)}\cos{\alpha},\;y=a\cos{(n\alpha)}\sin{\alpha}$ the formula (\ref{eq36}) can also be established by the methods of \cite{Aghamohammadi_2024} with due attention to the limits of integration over $\alpha$.

Interestingly enough, bumblebees and honeybees can sense the presence of weak electric fields around negatively charged flowers, discriminate between electric fields with different radial geometries, and use this information to assess floral reward. It is thought that electroreception in both bees and flowers, as well as electrostatic interactions between a positively charged bee and a negatively charged flower, may play a role in pollination \cite{Clarke_2017,England_2022}. Due to its simplicity, \ref{eq2} can be used to model electric potential in experiments studying this phenomenon where and when flat geometry prevails. In general case, both \ref{eq1} and \ref{eq1DPRD} can be calculated with almost equal ease on modern computers. However, having two different expressions for the same potential is useful to avoid introducing an accidental error in the computer program that calculates the potential. 

\section*{Declaration of competing interest}
The authors declare that they have no known competing financial interests or personal relationships that could have appeared to influence the work reported in this paper. 

\section*{Acknowledgements}
We would like to thank Max Alexeev and Martin Nicholson for helpful ideas on calculating the angular integral (\ref{eq19}) that they shared on \emph{Mathoverflow}. We appreciate the helpful correspondence from Amir Agamohammadi. The reviewers' comments helped improve the manuscript.

\appendix
\section{Proofs of the mathematical identities used}
First recall the generating function of the Legendre polynomials \cite{Askey_1999}
\begin{equation}
\frac{1}{\sqrt{1-2tx+t^2}}=\sum_{n=0}^\infty P_n(x)t^n.
\label{eq3}
\end{equation}
If we take $x=\cos{\theta}=\frac{1}{2}\left (e^{i\theta}+e^{-i\theta}\right)$, then
\begin{equation}
1-2tx+t^2=(t-e^{i\theta})(t-e^{-i\theta})=(1-te^{i\theta})(1-te^{-i\theta}).
\label{eq4}
\end{equation}
Now we can use the binomial theorem \cite{Askey_1999}
\begin{equation}
(1-x)^{-a}=\sum_{n=0}^\infty \frac{(a)_n}{n!}\,x^n,\;\;\;(a)_n=a(a+1)(a+2)\cdots(a+n-1),
\label{eq5}
\end{equation}
and get
\begin{equation}
\frac{1}{\sqrt{1-2tx+t^2}}=\sum_{m=0}^\infty\sum_{k=0}^\infty \frac{\Bigl (\frac{1}{2}\Bigr )_m\Bigl(\frac{1}{2}\Bigr)_k}{m!k!}\,t^{m+k}e^{i(m-k)\theta}.
\label{eq6}
\end{equation}
The double sum can be rearranged by introducing $n=m+k$ and noting that, for a fixed $n$, $k$ changes from $0$ to $n$ (since $m=n-k\ge 0$):
\begin{equation}
\frac{1}{\sqrt{1-2tx+t^2}}= \sum_{n=0}^\infty\sum_{k=0}^n
\frac{\Bigl(\frac{1}{2}\Bigr)_k\Bigl(\frac{1}{2}\Bigr)_{n-k}}{k!(n-k)!}\,e^{i(n-2k)\theta}\,t^{n}. 
\label{eq7}
\end{equation}
Comparing with (\ref{eq3}) and remembering that the left side is real, we get 
\begin{equation}
P_n(\cos{\theta})=\sum_{k=0}^n\frac{\Bigl(\frac{1}{2}\Bigr)_k\Bigl(\frac{1}{2}\Bigr)_{n-k}}{k!(n-k)!}\,\cos{(n-2k)\theta}.
\label{eq8}
\end{equation}
This relation is a special case of expression 6.4.11 for Gegenbauer polynomials from \cite{Askey_1999}.

The second identity is easier to establish. Using $\cos{\theta}=\frac{1}{2}(e^{i\theta}+e^{-i\theta})$ and the binomial theorem, we get
\begin{equation}
\cos^n{\theta}=\frac{1}{2^n}\sum_{k=0}^n \binom{n}{k}e^{i(n-2k)\theta}=
\frac{1}{2^n}\sum_{k=0}^n \binom{n}{k}\cos{(n-2k)\theta}.
\label{A_eq11}
\end{equation}

\bibliography{rose_disk}

\end{document}